\documentclass[manuscript]{aastex631}

\hypersetup{linkcolor=red,citecolor=green,filecolor=cyan,urlcolor=magenta}

\begin{document}

\title{Effect of viewing angle in Gamma-Ray Burst properties}
\correspondingauthor{Sreelakshmi P Chakyar}

\email{sreelakshmip.17@res.iist.ac.in}
\author{Sreelakshmi P Chakyar}
\affiliation{Indian Institute of Space Science and Technology, Trivandrum 695547, Kerala, India}
\author{Sarath Prabhavu J}
\affiliation{Indian Institute of Space Science and Technology, Trivandrum 695547, Kerala, India}
\affiliation{Aryabhatta Research Institute of Observational Sciences,\\ Manora Peak, Nainital-263001, Uttarakhand, India}

\author{Lekshmi Resmi}
\affiliation{Indian Institute of Space Science and Technology, Trivandrum 695547, Kerala, India}
\begin{abstract}
The empirical classification of Gamma-Ray Bursts (GRBs) is based on their distribution in the plane of burst duration and spectral hardness. Two distinct distributions, long-soft and short-hard bursts, are observed in this plane, forming the basis for the long and short classification scheme. Traditionally, this scheme was mapped to two different GRB progenitor classes. However, several recent bursts have challenged this mapping. This work investigates how an observer's viewing angle relative to the jet axis influences the duration-hardness plane. We simulate single-pulse GRBs using an optically and geometrically thin homogeneous top-hat jet model. Bursts are simulated with an isotropic viewing angle distribution, and we calculate the pulse duration and spectral hardness corresponding to \textit{FERMI} Gamma-Ray Burst Monitor (GBM) energy bands. The viewing angle significantly impacts spectral hardness for our assumed broken power-law spectra, while its effect on duration is less pronounced. Our analysis indicates that soft and low-luminous bursts are likely off-axis events. It is possible that some of the fast X-ray transients and X-ray rich GRBs observed by the Einstein Probe and the Space Variable Objects Monitor (SVOM) missions originate from off-axis jets.
\end{abstract}
\section{Introduction} \label{sec:intro}
Gamma-ray bursts (GRBs) are classified as long and short bursts based on the observed bimodal distribution in their duration. The spectral hardness of the $\gamma$-ray pulse is found to be correlated broadly with duration, leading to a classification of GRBs as long-soft and short-hard bursts \citep{chryssa}. 

Traditionally, these two classes are directly associated with two distinct progenitor systems where collapsing massive stars lead to long-soft bursts and merging neutron star binaries lead to short-hard bursts. Ideally, one considers that a longer-lasting accretion disk formed by the envelope of the collapsing star can lead to a longer duration burst \citep{1993ApJ...405..273W, 1998ApJ...494L..45P} while merging compact objects with a disk of much smaller mass can lead to shorter bursts \citep{progenitor1, 1992ApJ...395L..83N, sGRBprogenior2, 1999A&A...341..499R}. However, several recent bursts have challenged this hypothesis \citep{Zhang, Xu, 2021NatAs...5..911Z, 2022ApJ...931L..23L, 2022Natur.612..228T, 2022Natur.612..223R, grb211211a, grb211227a, grb230307a, burstchallenge}. 

Studies have indicated that the dividing line between long and short duration GRBs is sensitive to factors such as the detector energy band and burst redshift \citep{Bingshortlong1, Bingshortlong2, Qin_2013, tarnoposki_2015}, also see \citep{zhangchoi}. Relativistic effects can also influence the observed duration and hardness, which is used to explain X-ray flashes \citep{XRF_offGRB, xrf_030723, XRF080330, XRF020903_2} and low luminosity GRBs such as GRB 980425 and GRB 031206 \citep{lowlum_031203, lowlum_980425, lowlum1, low_lum_3}. 

The ultra-relativistic bulk velocities of GRB jets cause significant Doppler deboosting of observed flux when directed towards the observer. If there is considerable misalignment between the jet axis and the observer's viewing angle, relativistic effects reduce the brightness, leading to non-detection unless the burst is highly luminous or nearby. Therefore, most bursts detected are nearly on-axis. However, the probability of detecting an off-axis burst is not negligible if it is energetic, nearby, or both. For example, by modelling a large sample of X-ray afterglows \citet{ryan_off} inferred that a substantial fraction of bursts are not viewed along the jet axis. To explain the afterglow of the short GRB 140903A \cite{troja_off} invoked an off-axis GRB model. The presence of relativistic effects in such cases can also change the prompt emission properties, such as the burst duration, hardness, and fluence from what it would have been if observed closer to the jet axis \citep{salafialike1, salafialike3, salafialike2, salafialike4, 10.1093/mnras/stw1549}.  

An example of this is GRB170817A, associated with the binary neutron star merger GW170817. The $\gamma$-ray pulse was short (2.0$\pm0.5$), soft ($\sim$ 0.7), and faint ((1.4$\pm$0.3) $\times$ 10$^{-7}$erg cm$^{-2}$) corresponding to an isotropic equivalent energy of 4.17$\times$10$^{46}$ erg \citep{170817_detection}. Afterglow observations revealed that it was observed at $\sim 30^{\circ}$ off-axis angle \citep{margutti17, troja17, 2018ApJ...867...57R, lazzati18, lamb_2019}. 
This confirmed the potential presence of off-axis events in the population of GRBs detected so far. Motivated by this, several authors searched for and reported sub-populations of GBM bursts showing similar properties as GRB170817 \citep{grb150101b_1, grb150101b_2, gbmlike_170817_1}.

This paper explores how viewing angle affects the spectrum and light curve of a single pulse GRB. We assume an optically and geometrically thin relativistic shell with co-moving frame emissivity that decays exponentially over time \citep{Woods_1999}. We assume a broken power-law shape for the intrinsic spectrum of the pulse. We simulate the duration-hardness plane by varying the parameters of the pulse, such as the characteristic decay time-scale of co-moving frame emissivity, intrinsic spectrum, and the Lorentz factor of the jet. We examine how an off-axis GRB could manifest in the observed population and draw similarities between our simulated population and some recent bursts. We assume a uniform tophat jet in our calculations. However, there is growing observational evidence that both long and short GRBs have complex jet structures \citep{GRB150101B2, oconnor, strjet2}. In structured jets, where the bulk Lorentz factor and energy per solid angle gradually decrease away from the jet axis, the Doppler de-boosting is less severe, and the variation in burst properties with viewing angle can differ noticeably.

In Section \ref{sec:style}, we explain the theoretical model. In Section \ref{sec:spectrum}, we present the specturm and light curve of a single pulse off-axis GRB and in Section \ref{sec:sim} we discuss the simulated population of GRBs. Finally, we summarise our work in Section \ref{sec:summary}. 

\section{Calculation of prompt emission flux}\label{sec:style}
In this section, we describe the model used for calculating the prompt emission flux from a relativistic jet. We follow the formulation developed by \citet{Woods_1999} for a uniform top-hat jet. 

\begin{figure}[h!]
    \centering
    \begin{tabular}{cc}
    \includegraphics[width=9cm]{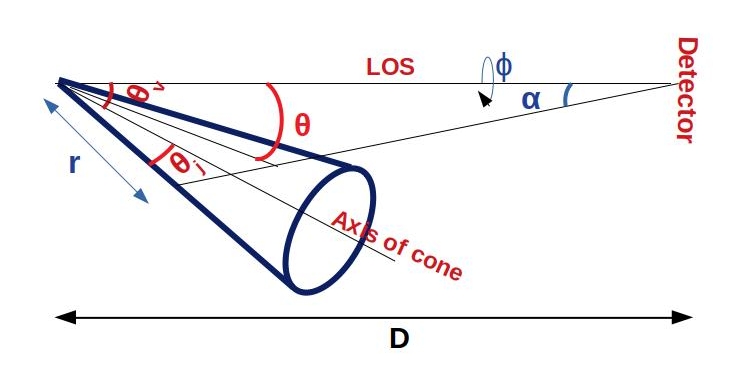}     & 
    \includegraphics[width=8cm]{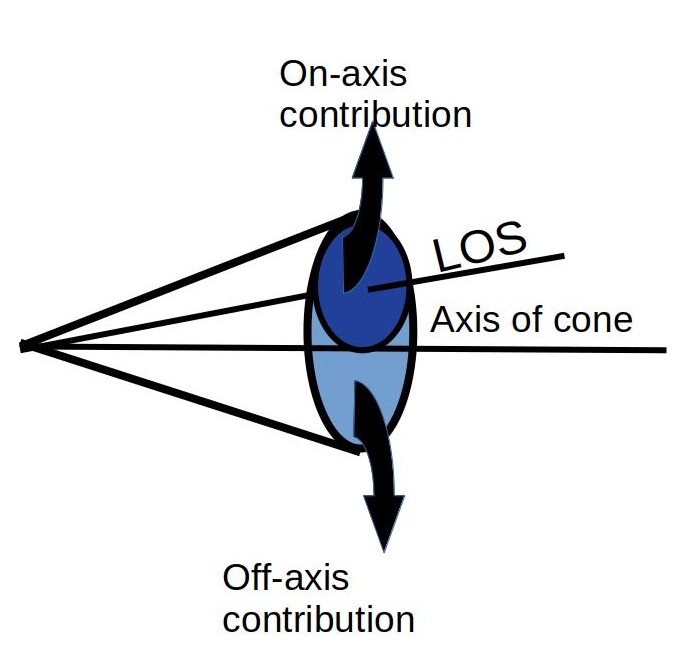}     \\ 
    \end{tabular}
    \caption{Left panel: Geometry of the jet and observer used in the formalism adapted form \cite{Woods_1999} with permission. Right panel: Illustration of an intermediate off-axis event for which the flux is calculated as the sum of an on-axis part and an extreme off-axis part.}
    \label{Fig:1}
\end{figure}

The flux $F_{\nu}(T)$ observed at a frequency $\nu$ at time $T$ can be expressed as,
\begin{equation}
 F_{\nu} (T) =\int{I_{\nu}(\alpha,\phi)d\Omega}=\int^{2\pi}_{0}{d\phi}\int_{0}^{\pi}{d\alpha\sin \alpha \cos \alpha \; I_{\nu}(\alpha,\phi)},
\label{eq:1}
\end{equation}
where $I_{\nu} (\alpha, \phi)$ is the intensity of the ray reaching the detector in direction $(\alpha, \phi)$ w.r.t the line of sight. The jet has a half-opening angle of $\theta_j$ while $\theta_v$ represents the angle between the observer's line of sight and the jet axis. See Figure \ref{Fig:1} (left) adapted from \citet{Woods_1999} with permission.
%
%

Considering the emitting plasma to be optically thin, the intensity can be expressed as an integral of the emissivity $j_{\nu} (t, r, \theta, \phi)${\footnote{Emissivity is defined as the specific power emitted per unit time by unit volume of the plasma per unit solid angle.}} over the line element $ds$ within the material along the line of sight to the observer. Here, $t$ is the time of emission of a photon, $r$ is the distance from the centre of the explosion, and $\theta$ and $\phi$ are the polar and azimuth angles w.r.t the line of sight to the observer. The progenitor's and observer's frames are at rest concerning each other and are related through light travel time effects. Hence, the emitted time $t$ and arrival time $T$ of photons are related by $T = t-\frac{r \mu}{c}$, where $\mu = \cos{\theta}$. That is, photons emitted at a higher latitude (larger $\theta$) w.r.t the observer will arrive with a delay compared to the photons emitted from angles close to the line of sight ($\theta \sim 0$).
 
The line element $ds$ can be approximated as $d\mu \frac{\alpha D}{(1-\mu^2)^{3/2}}$ for astrophysical distances \citep{Woods_1999}. Therefore, Equation \ref{eq:1} can be rewritten using the Lorentz invariance of $j_\nu/\nu^2$ as,
\begin{equation}
 F_{\nu} (T)=D\int^{2\pi}_{0}{d\phi}\int_{0}^{\alpha_{m}}{d\alpha \alpha^2 \int^{1}_{-1} {d\mu}\frac{j'_{\nu^{\prime}}(r, \theta, \phi, t)}{(1-\mu^2)^{3/2}}} (\nu/\nu^\prime)^2,
\label{eq:2}
\end{equation}
where $\alpha_m = L/D$. 

{\subsection{Pulse profile and jet structure}} \label{subsec:shape}
The emissivity function encapsulates the pulse profile, including its spectral and temporal characteristics and the jet structure. Following \citet{Woods_1999}, the jet co-moving frame emissivity $j_{\nu^{\prime}}^{\prime}$ is assumed to be,
\begin{equation}
 j'_{\nu^{\prime}}(r,t)=A(t) \, \delta(r-\beta ct) \Pi(\theta) \kappa(\nu^{\prime}). 
 \label{eq:3}
\end{equation}
The $\delta$-function imposes the assumption of the geometrically thin shell. At a given time $t$, photons are emitted only from an infinitesimally thin shell of $r = \beta c t$. In terms of the observed time $T$, this can be written as $\delta[T-\frac{r(1-\beta \mu)}{\beta c}]$ leading to the formation of an egg-shaped equal arrival time surface from where the observer receives photons at a given $T$ \citep{rees_eot}. As $\mu, \alpha,$ and $r$  are related, this $\delta$-function can simplify the integration over $\alpha$ in Equation \ref{eq:2}. 

The temporal part of the emissivity function assumes an exponential decay of the pulse, $A(t)=A_{0}e^{-t/\tau}$ where $A_0$ is a normalization constant and $\tau$ is the characteristic decay timescale. The shell expands with a constant velocity $\beta$. At the same time, the pulse decays exponentially. As a result, the final flux increases till a maximum and falls exponentially afterwards.

$\kappa(\nu^{\prime})$ is the spectral shape function defined in the co-moving frame where  $\nu^{\prime}$ is the co-moving frame frequency. In this paper, we consider a broken power-law spectrum, as given below.
\begin{equation}
\kappa_{\rm bpl\textit{\textit{}}}(\nu^{\prime})=
\left\{
        \begin{array}{cc}
        0,     & \nu^{\prime} <1\times10^{-4} \; \; {\rm keV} \\
        (\nu^{\prime}/\nu_b^{\prime})^{-m_1},     & 1\times10^{-4} \; \; {\rm keV} \leq \nu^{\prime}\leq \nu_b^{\prime} \\
        (\nu^{\prime}/\nu_b^{\prime})^{-m_2},   &  \nu^{\prime} > \nu_b^{\prime},
        \end{array}
\right.
\label{eq:4}
\end{equation}
where $m_1$ and $m_2$ are power-law indices and $\nu_b^{\prime}$ is the break frequency expressed in units of $\rm keV$ throughout this paper. With the current instruments, we do not have information on the GRB prompt emission spectrum below $1$~keV (except some bursts detected at the end of the prompt phase by {\textit{Swift}} XRT in $\sim 0.3-10$~keV. For a Lorentz factor of $\le 1000$, $1$~keV in the observer frame will correspond to $\nu^{\prime} \le$ $\sim 10^{-4}$~keV for an on-axis observer.  Therefore, we have applied a spectral cut-off below $\nu^{\prime} = 10^{-4}$~keV.  For an off-axis observer, this limiting value will be lower. No observable effects are introduced within the 10-1000 keV band under consideration due to this cut-off.

$\Pi(\theta)$ imposes the jet structure which depends on the observer's viewing angle. For example, for an on-axis observer ($\theta_v = 0$), $\Pi(\theta) = H(\theta_{j}-\theta)$, the Heaviside step function. The functional form of $\Pi(\theta)$ is more complex for an off-axis observer (see below).

\subsection{Flux received by an on-axis observer}\label{subsec:onaxis}
For an on-axis observer ($\theta_v=0$), for whom the jet axis coincides with the line of sight, the jet is symmetric in both $\theta$ and $\phi$. Hence $\Pi(\theta)$ becomes the Heaviside step function $H(\theta_{j}-\theta)$.

Due to axial symmetry, the integral over $\phi$ is $2 \pi$. The Doppler effect relates the observed and emitted frequencies through the component of the velocity of the emitting material along the line of sight. If the jet expands radially, this leads to $\nu/\nu^{\prime} = \frac{1}{\gamma(1-\beta \mu)}$. For a given observed frequency, $\nu^{\prime}$ varies across the jet as $\theta$ varies based on this formula. Using this, \citet{Woods_1999} converts the integration over $\mu$ to an integration over $\nu^{\prime}$. The limits of the $\mu$ integration (i.e., of the $\nu^{\prime}$ integration after changing variables) depend on the observer's viewing angle. 

Finally, Equation \ref{eq:1} can be rewritten as,
\begin{equation}
 F_{\nu}(T)=2\pi\beta\gamma^{2}\left(\frac{cT}{D}\right)^2\nu^4\int^{\nu\gamma(1-\beta\cos\theta_{j})}_{\nu\gamma(1-\beta)}f(\nu^{\prime})\, A_0 e^{-\left(\frac{\nu\gamma}{\nu^{\prime}}\frac{T}{\tau}\right)} \, \frac{d\nu^{\prime}}{{\nu^{\prime}}^5},
 \label{eq:5}
\end{equation}\\
for the flux received by an on-axis observer. The $\delta$ function corresponding to the thin shell assumption is used to rewrite $A(t)$ in terms of $T$. Additionally, as noted in the previous subsection, the $\delta$ function simplifies the $\alpha$ integration.

\subsection{Flux received by an off-axis observer}\label{subsec:offaxis}
For an off-axis observer, $\theta_v \neq 0$. The off-axis angles can be divided into two types: moderately off-axis ($\theta_v < \theta_{j}$ and extreme off-axis ($\theta_v >\theta_{j}$) angles. 

For an extreme off-axis observer, the emitting material is confined to a cone with a boundary defined by $\cos\theta\cos\theta_{v}+\sin\theta_{v}\sin\theta\cos\phi=\cos\theta_{j}$ \citep{Woods_1999}. This leads the jet profile to a combination of two step functions in $\theta$ and $\phi$, and the emissivity can be written in this case as, 
\begin{equation}
 j'_{\nu^{\prime}}(r,t, \theta, \phi)=A(t)\delta(r-\beta ct)H(\theta_{j}-| \theta - \theta_{v}|)H\left[\cos\phi-\left(\frac{\cos\theta_{j}-\cos\theta\cos\theta_{v}}{\sin\theta_{v}\sin\theta}\right)\right]f(\nu^{\prime}).
 \label{eq:6}
\end{equation}

This essentially changes the limits of the $\nu^{\prime}$ integration as given in the below expression. Finally $F_{\nu}(T)$ in the extreme off-axis case can be written as,
\begin{equation}
 F_{\nu}(T)=2\Delta\phi\beta\gamma^{2}\left( \frac{cT}{D} \right)^2\nu^4\int^{\nu\gamma(1-\beta\cos(\theta_{v}-\theta_{j}))}_{\nu\gamma(1-\beta\cos(\theta_{v}+\theta_{j}))}f(\nu^{\prime})\, A_0 e^{-\frac{\nu\gamma}{\nu^{\prime}}\frac{T}{\tau}}\,\frac{d\nu^{\prime}}{{\nu^{\prime}}^5}.
 \label{eq:7}
\end{equation}
The moderately off-axis case can be considered a sum of two contributions: a symmetric part around the line of sight and an extreme off-axis wedge-shaped part (see Figure \ref{Fig:1} (right)) with appropriate limits of $\nu^{\prime}$ integration. See \citet{Woods_1999} for more details.

\section{Impact of Viewing Angle on Spectrum and Light Curve}\label{sec:spectrum}
%
%
\begin{figure}
    \centering
    \includegraphics[width=0.9\textwidth]{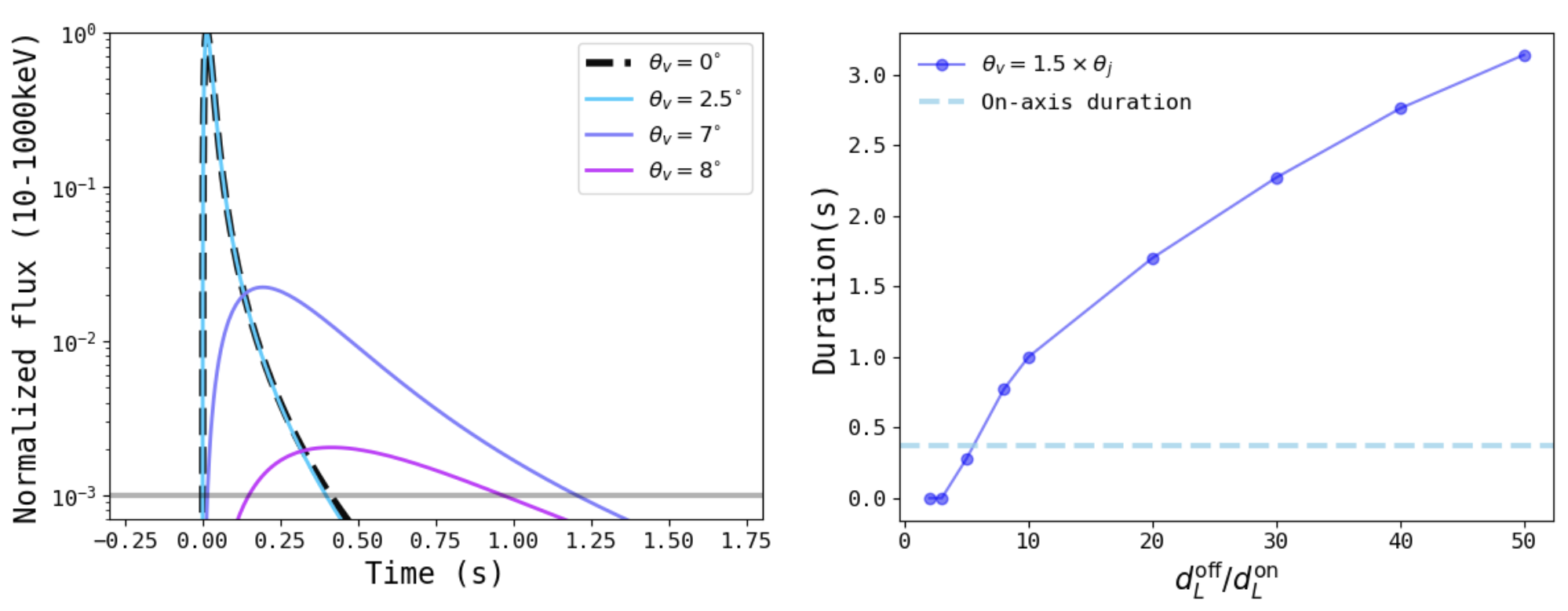}
    \caption{The light curve for different viewing angles (left). The y-axis is normalized. The parameters used are $\theta_j = 5^{\deg}, \gamma=100, \tau = 100$~s, $m_1=0.1, m_2 = 1.2, $ and $\nu_b^{\prime} = 1.5$~keV. For $\theta_v < \theta_j$, the flux is nearly the same as the on-axis ($\theta_v =0$) one. The pulse stretches due to the delay in the arrival of high-latitude photons. Duration of identical bursts detected at different distances as $\theta_v$ varies (right). In the case presented here, the on-axis duration is $0.4$ seconds. We calculate the duration for $\theta_v = 7.5^{\circ}$ ($1.5 \times \theta_j$) by bringing the burst closer by $0.5$ to $0.02 \times$ the on-axis distance. We find that the burst can move from short to long duration w.r.t the boundary of the two classes for GBM (2 seconds).}
    \label{Fig:2}
\end{figure}

\begin{figure}[h!]
    \centering
    \begin{tabular}{cc}
     \includegraphics[width=9cm]{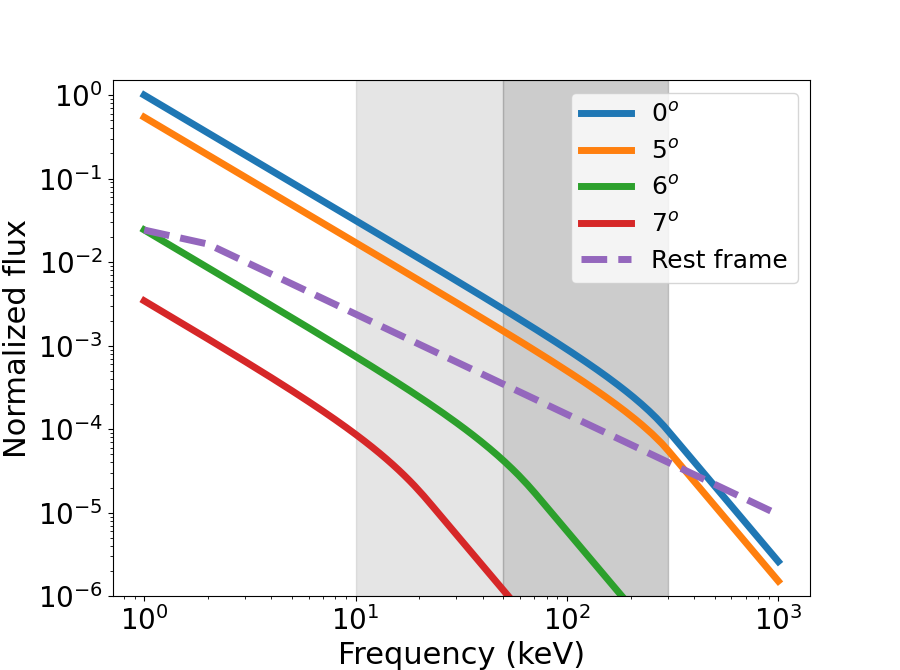}    &  
      \includegraphics[width=9cm]{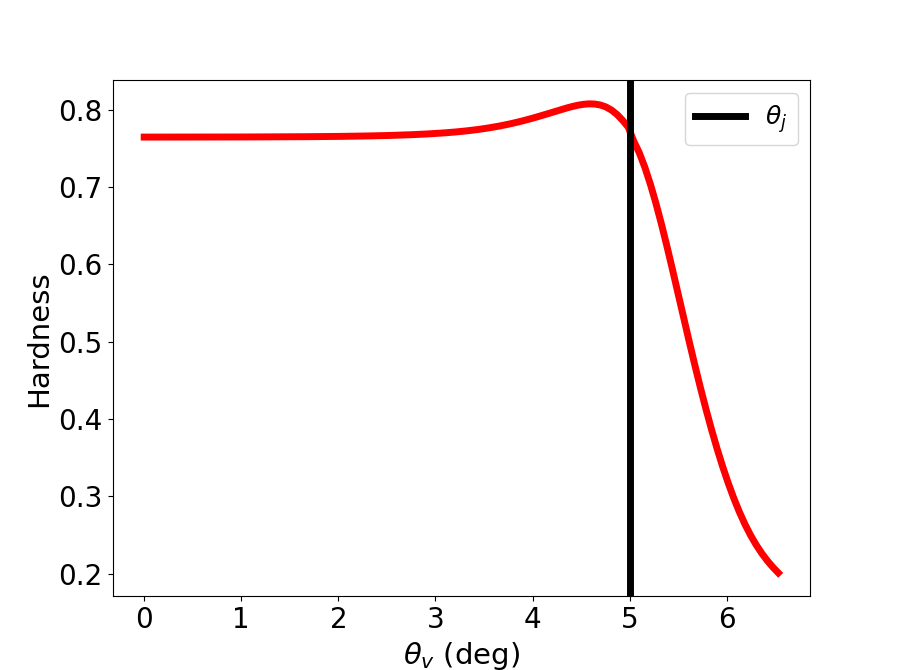} \\   
    \end{tabular}
    \caption{Change in the spectrum (left) as a function of viewing angle. The Y-axis is normalized. The dashed line represents the co-moving frame spectrum (scaled up). The parameters used are $\gamma=$100, $\tau=1000$~s, $E_{\gamma, \rm iso}=10^{51}$~erg, $\theta_{j}=5^{\circ}$, $d_L=200$~Mpc, m$_{1}$=0.5, m$_{2}$=2, $\nu_b^{\prime}=1.5$~keV. Shaded regions correspond to the energy bands used to calculate the hardness presented in the right panel. As $\nu^{\prime}_b$ shifts to lower values, the burst becomes softer. The variation of hardness (right) ($\frac{F(50-300)}{F(10-50)}$) with viewing angle. Parameters are the same as the left panel except $m_{1}$=0.1 and m$_{2}$=1.2.}
    \label{Fig:3}
\end{figure}
In this section, we explore the properties of the light curve and the time-integrated spectrum of the single-pulse GRB described above. 

We generate the light curves $F(T) = \int_{10 \rm keV}^{1000 \rm keV} d\nu F_{\nu}(T)$ for a range of viewing angles of the observer. We consider the $\textit{Fermi}$ GBM bandpass of $10-1000$~keV for frequency integration. Both duration and hardness start to differ only when $\theta_v \gtrsim \theta_j$ for a uniform top hat jet.

As the viewing angle increases and exceeds the half-opening angle of the jet, the light curve begins to stretch considerably. This is due to the increase in the arrival time, $T=t-r\mu/c$ of photons from higher latitudes ($\theta$). As $\theta_v$ increases, more and more photons arrive from higher latitudes.  As a result, the pulse broadens, and the peak of the light curve shifts to larger $T$ values. Figure \ref{Fig:2} (left) presents the normalized pulse shape. The y-axis is normalized as the figure is meant to focus on the change in shape alone and, hence, does not reflect the reduction in the flux. 

It is reasonable to suggest that a gradual increase in photon counts could delay the detector’s trigger time. If this is the case, part of the $1.7$ second delay observed between the merger and the GBM trigger of GRB 170817A may be due to the misalignment of the jet. Also see \citep{timedelay2} and \citep{timedelay}, who attributes the delay time of $1.7$ second, which is also comparable to the duration of the burst, to the time taken for the jet to propagate to the dissipation site.

For calculating the duration, we consider a threshold flux of $0.5$ photons $\rm cm^{-2}$ $ \rm s^{-1}$ corresponding to GBM. We define duration as the time period during which the pulse remains above the threshold. In this study, we make the critical assumption that the duration thus defined is a measure of $T_{90}$, the interval during which the fluence increases from 5\% to 90\% of the total fluence. If peak flux falls below the threshold, the pulse is considered undetected.

In principle, the broadening can increase the duration \citep{10.1093/mnras/stw1549}. Still, since the peak flux reduces, the width of the pulse above the instrument threshold essentially reduces the duration of the burst \citep{2021NatAs...5..911Z}.

However, nearby bursts of larger viewing angles can have a longer duration than their identical twins at higher distances. In Figure \ref{Fig:2} (right), we present the change in duration of a burst viewed at $\theta_v = 1.5 \times \theta_j$ at different distances. In another example with the same $\theta_v$, we observe that a burst with an on-axis duration of $\sim 4$~s has a duration of $0.5$ seconds when brought closer to us at a distance of $1/7$ times the original distance. Although the case of GRB170817 is not directly relevant here due to its Gaussian jet structure, such nearby bursts viewed highly off-axis can be of longer duration than their on-axis counterparts observed from a greater distance.

Next, we calculate the time-averaged spectrum of the simulated GRB. In Figure \ref{Fig:3} (left), we present the spectrum corresponding to four different viewing angles for a jet of opening angle of $5^{\circ}$. The other parameters are $\gamma=$100 and $\tau=$1000s. The spectral indices $m_1$ and $m_2$ are $0.1$ and $1.2$ respectively. The break frequency $\nu_b^{\prime} = 1.5$~keV. For an off-axis observer, the photons will be less blue-shifted than an on-axis observer, leading to a leftward shift of the spectral break, as seen in the figure. 

The spectral hardness is defined as the ratio of photon flux in the high energy band ($50-300$~keV) to the same in the low energy band ($10-50$~keV). The hardness drops as $\nu_b$ (equivalent to $E_{\rm peak}$) decreases for higher viewing angles after slightly increasing around $\theta_v \sim \theta_j$. In Figure \ref{Fig:3} (right), we present the variation in hardness for the same burst with respect to the viewing angle, ranging from $0^{\circ}$ to $7^{\circ}$. We can see a considerable variation in hardness for this set of parameters. The hardness for the broken power-law spectrum considered here is primarily determined by $m_2$ and ${\nu_b}^{\prime}$. The degree of hardness variation as a function of $\theta_v$ is mainly determined by the changing position of ${\nu_b}^{\prime}$ inside the detector band. 

\section{Simulation of the burst population}\label{sec:sim}
In the previous section, we demonstrated how the pulse's temporal and spectral shape changes as a function of the viewing angle. A larger $\theta_v$ leads to a broader pulse, implying a longer duration technically, but in effect, the Doppler deboosting will reduce the duration in almost all cases. The spectrum becomes softer for large values of $\theta_v$. In this section, we simulate a population of bursts for a random choice of parameters. The intrinsic parameters deciding the burst properties are $\Theta=(\gamma, \theta_{j}, \tau, m_1, m_2, \nu_b, E_{\gamma, \rm iso}, d_L, \theta_v)$.

Before proceeding, we need to use the on-axis fluence to express the normalization constant $A_0$ in terms of the isotropic equivalent energy $E_{\gamma \rm iso}$ and the luminosity distance $d_L$. i.e, $E_{iso} = 4 \pi d_L^{2}\int_{10}^{1000} S_{\nu} (\theta_v = 0) d \nu$, where S$_{\nu} (\theta_v =0)$ is the on-axis fluence at $\nu$, and is calculated by integrating Equation \ref{eq:5} over frequency \citep{Woods_1999}. 
\begin{equation}
    S_{\nu} (\theta_v=0) =\int_{0}^{\infty}F_{\nu}(T)dT=\frac{4\pi\beta A_{0}\tau\nu}{\gamma}\left(\frac{c\tau}{D}\right)^{2}\int_{\nu\gamma(1-\beta)}^{\nu\gamma(1-\beta\cos\theta_{j})}f(\nu^{'})\frac{d\nu^{'}}{\nu^{'2}}.
    \label{eq:8}
\end{equation}
A caveat to remember is that this integration uses $T$ from $0$ to $\infty$ while the real on-axis fluence will be measured for the width of the detected pulse. The difference, however, is negligible as most of the flux is received near the pulse peak. 

At cosmological distances, the k-correction will lead to spectral changes. This paper restricts the distance to $500$~Mpc ($z \sim 0.1$) and ignores k-correction.

\subsection{The on-axis pulse}\label{subsec:OAP}
%
\begin{figure}
    \centering
    \includegraphics[width=1.1\textwidth]{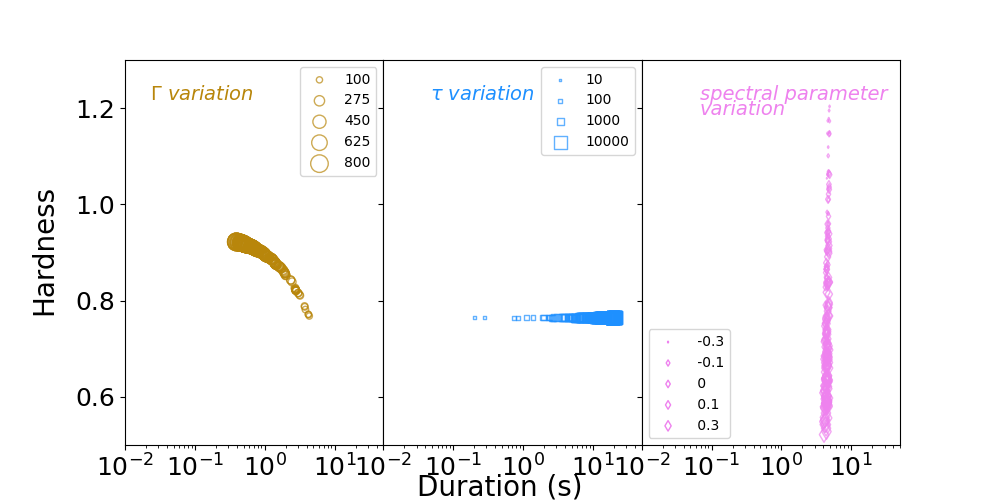}
    \caption{The duration- hardness plane for on-axis GRBs for varying the bulk Lorentz factor $\gamma$ (left), pulse decay timescale $\tau$ (middle), and spectral parameters (right). The remaining parameters are kept constant (see section \ref{subsec:OAP} for the value used when a parameter is fixed). The marker sizes from left to right are $\gamma$, $\tau$, and $m_1$, respectively. As $\gamma$ increases the pulse narrows down due to decreased arrival time of photons. The increase in hardness is due to increase in $\nu_b$. Increase in $\tau$ leads to an increase in duration, and $\tau$ does does not change hardness. Similarly a higher value of the the high energy spectral index $m_2$ increases the spectral slope and decreases hardness.} 
    \label{Fig:4}
\end{figure}
First, we examine how changes in $\gamma$, $\tau$, and the spectral parameters individually affect the hardness and duration (see Figure \ref{Fig:4}). During these runs, we keep all other parameters fixed. When fixed, the parameter values are $\Theta = (100., 5^{\circ}, 1000 {\rm s}, 0.1, 1.2, 1.5{\rm keV}, 10^{51}{\rm erg}, 200\, {\rm Mpc}, 0^{\circ})$. We did not consider the effect of $\theta_j$ in these runs because the ratio $\theta_v/\theta_j$ is the key parameter in deciding the properties.

The spectral parameters influence hardness significantly, but the influence on duration is secondary and less significant. We fix other parameters and vary spectral parameters in the range $ -0.3 < m_1 < 0.3, 1 < m_2 < 1.2$, and $ 0.5 < \nu_b^{\prime} < 1.5 $~keV.  To decide this range, We generate on-axis spectrum by varying these three parameters and fit it with the broken power-law function used in the GBM spectral catalogue \citep{Gruber_2014}. We chose the range of $m_1, m_2$ and $\nu_b^{\prime}$ such that the fitted parameters of the resultant on-axis spectrum are in agreement with the range of the observed parameters reported by \citet{Gruber_2014}.  See Figure $\ref{Fig:4}$ (right).

Both $\gamma$ and $\tau$ affect the duration due to their presence in the exponent of $F(T)$, $\exp^{-\gamma \nu T /\nu^{\prime} \tau}$. The pulse peak, that is, the maximum of the lightcurve ($\ \frac {d}{dT} \exp^{-\gamma \nu T /\nu^{\prime} \tau}|T_{p} = 0$) is proportional to $\tau/\gamma$. When either $\gamma$ increases, or $\tau$ decreases, the function $\exp^{-\frac{\gamma \, T}{\tau} }$ will fall faster leading to a shorter pulse. See Figure \ref{Fig:4} (left).

We vary $\gamma$ uniformly from $100 - 800$. The range of $\gamma$ is chosen based on the typical values found in the literature \citep{2009A&A...498..671X, 2011ApJ...726...89Z, 2017MNRAS.465..811N}. To decide the range of $\tau$, we fix other parameters and examined the duration of the resultant on-axis pulse. We find that $10 {s} < \tau < 10000{s}$, the on-axis duration is between $\sim 0.1$ to $\sim 50$~seconds for the above set of parameters. Obviously, variation in $\tau$ does not affect hardness.  See the Figure \ref{Fig:4} (middle).

Having understood the extent of variation possible in the duration-hardness plane for the ranges chosen for the above three parameters, we aim to simulate a GRB population of random viewing angles. 
\subsection{Off-axis pulse}\label{subsec:offpulse}
To produce the final simulated population, we vary all parameters in $\Theta$ except $\theta_j$. The spectral parameters, $\gamma$ and $\tau$, are randomly sampled from uniform distributions within the ranges specified in the previous subsection. $E_{\gamma, \rm iso}$ and $d_L$ are sampled as $\log (E_{\gamma, \rm iso}) \sim \mathcal U (50,52)$ and $d_L^3 \sim \mathcal U (200^3, 500^3)$ respectively. The isotropic equivalent energy and luminosity distance do not change the pulse shape but the burst intensity, affecting the duration.  

We generate on-axis pulses for $1000$ realizations of these parameters. Next, we randomly assign a non-zero viewing angle to the same pulses, following $\cos{\theta_v} \sim \mathcal U(-1, \cos{8^{\circ}})$. As a result, each pulse has an off-axis duration/hardness and a corresponding on-axis duration/hardness. 

\begin{figure}
    \centering
    \begin{tabular}{cc}
     \includegraphics[width=9cm]{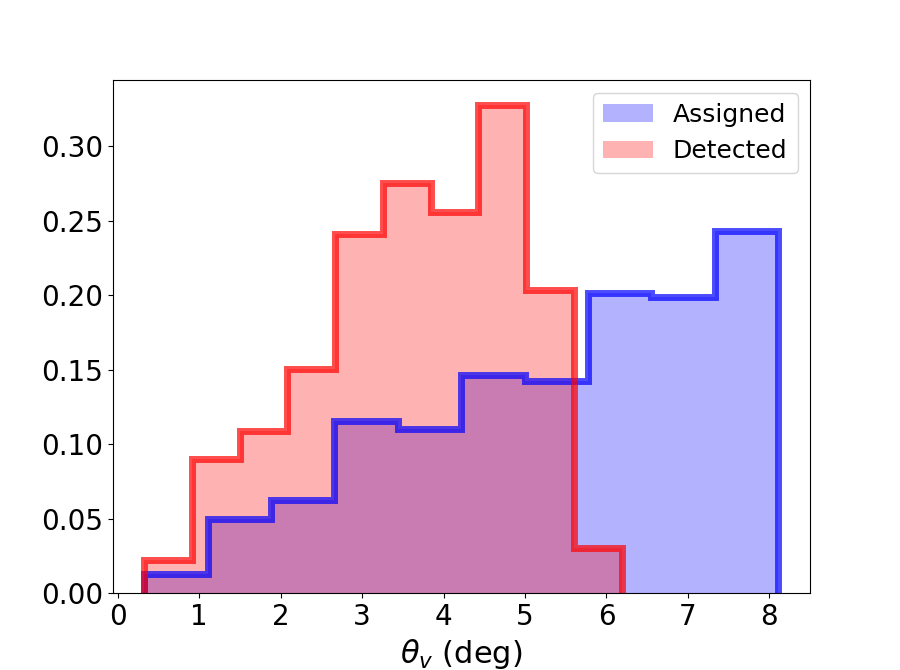}  &
      \includegraphics[width=9cm]{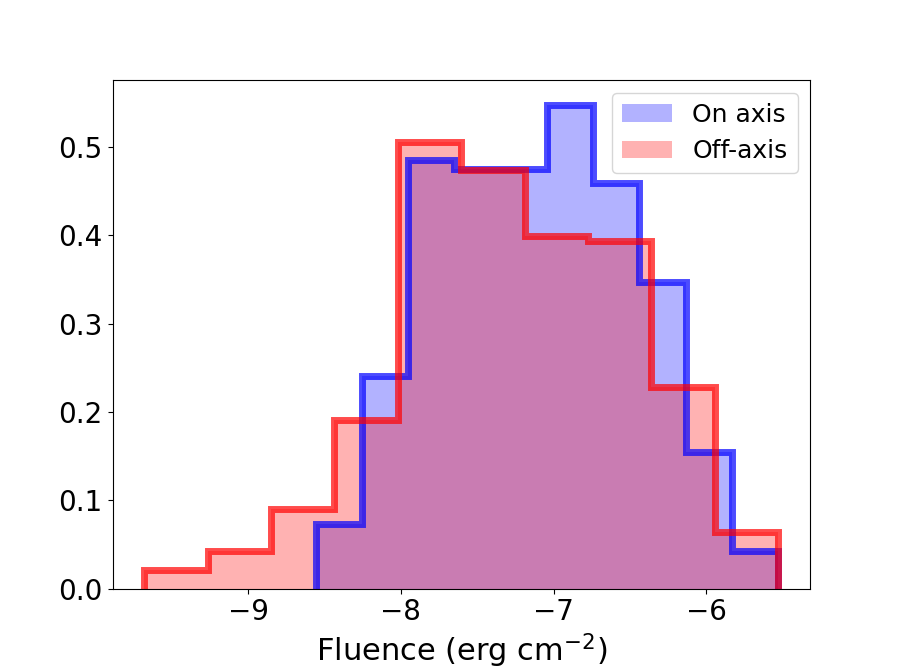} \\
    \end{tabular}
    \caption{The left panel presents the distribution of $\theta_v$ values of the simulated (blue) and detected (red) bursts. The right panel presents the distribution of the on-axis fluence (blue) and the off-axis fluence (red). }
    \label{Fig:5}
\end{figure}

We ran the simulation multiple times, and the results were consistent between runs. We restricted the viewing angle to $8^{\circ}$ as no bursts were detectable beyond this. In the one we present here, out of the $1000$ events, for $388$ one $\theta_{v}$ $\leq$ $\theta_{j}$ is assigned and for 612 events $\theta_{v}>\theta_{j}$ is assigned. Depending on $E_{\gamma, \rm iso}$ and $d_L$ values, some pulses obviously fall below the detection limit when $\theta_v$ increases. All events with $\theta_v \leq \theta_j$ are detected, and out of the $\theta_{v} > \theta_{j}$ events, $65$ events ($\sim 11$\%) are detected. When we changed $\theta_j = 3^{\circ}$ and kept $\theta_v < 8^{\circ}$, only $55$ out of the $865$ bursts ($\sim 6$\%) with  $\theta_{v} > \theta_{j}$ are detected. This shows that in the observed population of GRBs, extremely off-axis events are highly likely to be present. In the left panel of Figure \ref{Fig:5}, we present the histogram of the assigned $\theta_v$ values (blue) and the distribution of $\theta_v$ of the detected (red) population. The right panel of Figure \ref{Fig:5}, presents the observed fluence in both on- and off-axis populations. The off-axis events as a population have a lesser fluence. The distribution will differ if a broader duration of $E_{\gamma, \rm iso}$ and $d_L$ is used.

\begin{figure}
    \centering
    \begin{tabular}{cc}
    \includegraphics[width=10cm]{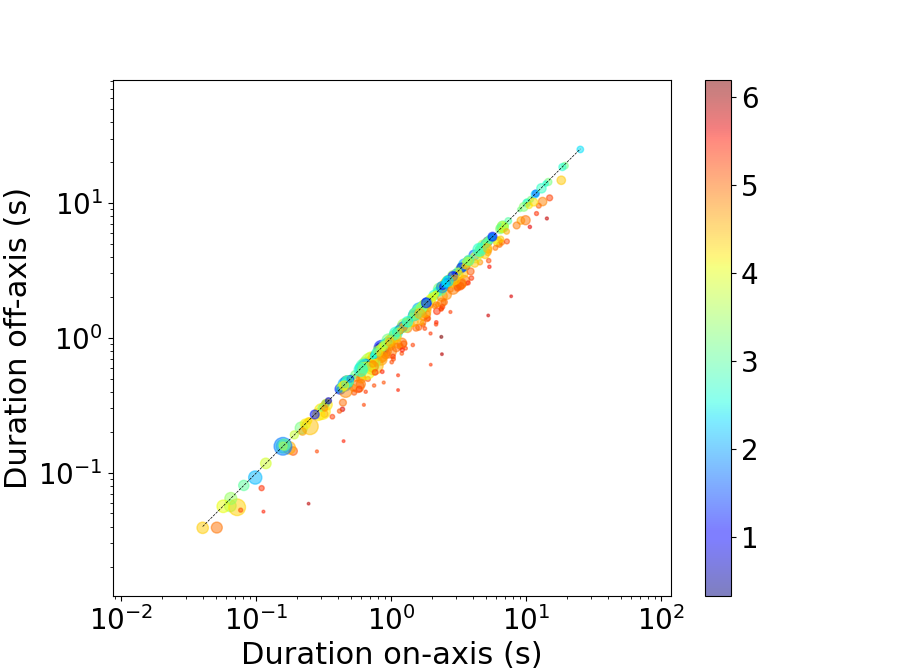} &
    \includegraphics[width=10cm]{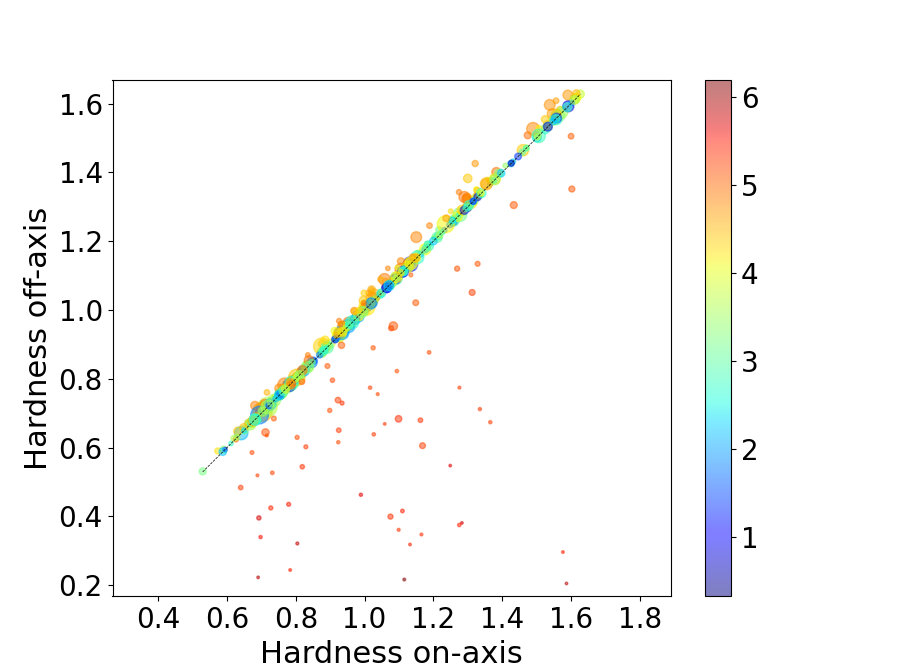} \\
    \end{tabular}
    \caption{Every burst is simulated for a $\theta_v = 0$ (on-axis) and a $\theta_v >0$ (off axis). The figure on the left (right) presents the on-axis vs off-axis duration (hardness) for the population. The figure is colour coded with the value of $\theta_{v}$ in degrees and the point sizes correspond to the peak flux. This figure illustrates that for the broken power-law spectrum with the range of parameters we have used there is a significant reduction in hardness for off-axis events while the range of durations of off-axis events are well within the on-axis ranges.}
    \label{Fig:6}
\end{figure}

In Figure \ref{Fig:6}, we present the on-axis and off-axis values of duration (left) and hardness (right) for the detected population. We see that the distribution of the on-axis duration extends from $\sim 0.3$ to $\sim 30$~seconds. All the off-axis bursts have a lesser duration. The variation in duration is up to $\sim 80$\%. If we have increased the range of $E_{\gamma, \rm iso}$ and $d_L$ in addition to $\theta_v$, we could have seen an increase in duration in some cases. 

Since, in our calculations, hardness is the highest for $\theta_v \sim \theta_j$, there are off-axis bursts with an increase in hardness. The hardness distribution extends to lower values for the full population than the on-axis population, making hardness a key property in identifying off-axis bursts. The change in hardness is highly sensitive to the $\kappa_{\nu^{\prime}}$, both its functional form and the parameters used. For a different set of parameters, particularly a lower value of $m_2$, the spread in hardness will be smaller.
%

\subsection{The hardness-duration plane for off-axis bursts}\label{subsec:HD}
\begin{figure}
    \centering
    \includegraphics[width=10cm]{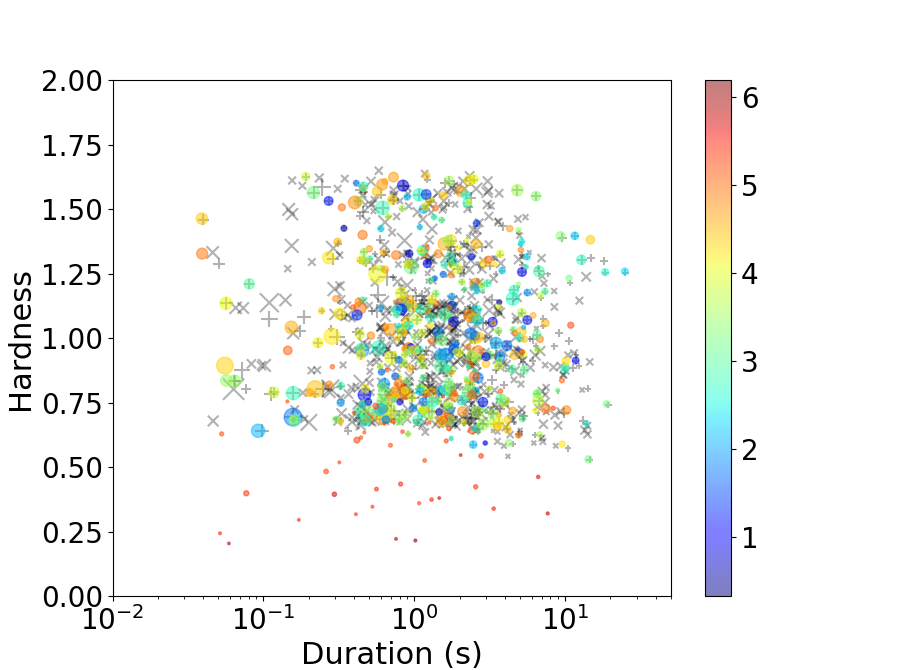}
    \caption{The duration-hardness distribution of the detected events from our simulation is shown. Gray symbols represent on-axis events ($\theta_v=0$). Among these, `+' sign indicates bursts that remain detectable when $\theta_v$ is made non-zero, while 'x' sign indicates bursts that become undetectable when their $\theta_v$ is increased from zero. Off-axis ($\theta_v >0$) bursts are colored according to $\theta_v$. In both cases, the symbol size varies with the peak flux}.
    \label{Fig:7}
\end{figure}

In Figure \ref{Fig:7}, we present the final duration-hardness plane for all the detected bursts (both on-axis and off-axis) in the simulated population. 

As described in the previous subsection, under the assumptions of our model and simulations, the change in hardness is more significant than the duration with respect to the viewing angle. The result can change depending on the spectral shape and parameters, which we will postpone until another paper. 

Our results show that fluence and hardness, not duration, are the key parameters to identifying off-axis bursts. Soft bursts with a lesser fluence have a high chance of being extreme off-axis events.

We aim to understand the changes resulting from higher viewing angles, rather than replicating an observed population. To reproduce the bimodality of the observed duration distribution, the distribution of $\tau$ would need to be changed, possibly to a bimodal one, and the spectral shape would need to be adjusted. Moreover, instead of the simplified duration we use, $T_{90}$, the time required to detect 90\% of the total counts, is to be calculated more rigorously for comparing with the observed population. 

Similarly, the observed hardness distribution is strongly sensitive to the assumed spectral shape and distribution of spectral parameters. The detected fraction of off-axis bursts is sensitive to the distribution of jet opening angle, bulk Lorentz factor, and luminosity function (here, the distribution of $E_{\gamma, \rm iso}$). We have assumed simple representative distributions for these in order to understand the effect of viewing angle on duration and hardness. Therefore, while spectral softness is a consequence of a high viewing angle, the large population of soft bursts need not correspond to the observable distribution. Due to these reasons, a comparison of Figure \ref{Fig:7} with the observed population is not appropriate. 

The low isotropic-equivalent energy and relatively softer spectrum of GRB 170817A, compared to the short GRB population, align with the predicted characteristics of off-axis bursts. GRB 150101B is considered a cosmological analogue of GRB 170817A due to the similarities in the optical and X-ray counterpart behaviour \citep{GRB150101B2}. Although the prompt emission of both bursts is phenomenologically similar, GRB 150101B is not as spectrally soft as GRB 170817A \citep{GRB150101B1}. It may be possible to explain these characteristics by a moderately off-axis angle. However, it is important to note that our model assumes a homogeneous top-hat jet, whereas both GRB 170817A and GRB 150101B are thought to be structured jets based on afterglow modeling.

While this manuscript was in preparation, the Einstein Probe (EP, \citeauthor{EP2} \citeyear{EP2}) started detecting Gamma-Ray Bursts with the Wide-Field X-ray Telescope (WXT, $0.5$ to $4$~keV) on board. This unprecedented soft X-ray view of the GRB prompt emission unveiled several mysteries \citep{EPdurdiff, EPresult3, EPresult5,  EPresult6, EPresult7}. In cases where there is simultaneous detection, it is found that the duration of the pulse in the WXT bands is longer than that in the GBM or BAT band \citep{EPdurdiff}.

Since the submission of this manuscript, two independent studies have been made available publicly where misaligned jets were invoked to explain X-ray transients similar to what EP has been detecting. \citet{EPresult1}, by using predominantly afterglow modelling, proposed that extra-galactic fast X-ray transients could originate from off-axis jets. \citet{EPresult7} invoked misaligned jets to explain the peculiar prompt emission characteristics of the EP GRBs. Their study also revealed that low fluence and soft spectra could originate from the misalignment of jets, a conclusion consistent with our findings.
At the same time, the ECLAIRS telescope on board the SVOM mission \citep{ SVOM2} also started reporting X-ray rich GRBs since its commissioning phase. 

\section{Summary and conclusion}\label{sec:summary}
In this paper, we explore how viewing angle can change the position of a burst in the hardness-duration plane used for the empirical classification of GRBs. We use a uniform top-hat jet model to simulate single-pulse GRBs. We assume an optically and geometrically thin shell with its emissivity decaying exponentially with time. The spectrum is assumed to be a broken power-law. The parameters under this model are the bulk Lorentz factor ($\gamma$), the decay timescale of the pulse in the co-moving frame ($\tau$), the half-opening angle of the jet ($\theta_j$), power-law indices ($m_1, m_2$) and co-moving frame break frequency ($\nu_b^{\prime}$) of the spectrum, the isotropic equivalent on-axis energy $E_{\gamma, \rm iso}$, and the luminosity distance ($d_L$). 

Due to relativistic effects, the peak flux reduces considerably for extreme off-axis ($\theta_v >\theta_j$) bursts, as expected. 
Intermediate off-axis $(\theta_v <\theta_j)$ events are not different from their on-axis counterparts. We find that a decent fraction of off-axis bursts can be detected for $50 < log_{10}(E_{\gamma, \rm iso}) < 52$ and $200 < d_L < 500$~Mpc. We find 11 (6) \% of extreme off-axis $(\theta_v > \theta_j)$ bursts are detected for $\theta_j =5^{\circ} (\theta_j=3^{\circ})$ for the range of parameters we have considered. 

The pulse stretches for large viewing angles due to increased photon arrival times from higher latitudes, yet the observed duration reduces due to reduced brightness. However, an off-axis nearby burst can appear longer than its identical twin observed on-axis at a higher redshift. At the same time, the twin of an on-axis long burst can appear as a short burst if observed at sufficiently large viewing angles. Due to these reasons, we find that duration is not useful for identifying potential off-axis bursts. 

The spectral hardness, defined as the ratio of fluences in $50-300$~keV to $10-50$~keV bands, drops for extreme off-axis angles. Therefore, we find that the characteristics of extreme off-axis bursts are soft spectra and low fluence. We would like to stress that the idea of soft and faint off-axis GRBs has been previously employed to explain the soft spectra of X-ray flashes \citep{ XRF_offGRB, xrt_theory_2} and low-luminosity GRBs \citep{LLgrb_theory_1}. 

Our analysis assumes a uniform top-hat jet. It is important to note that the properties of GRBs from misaligned jets are expected to be sensitive to the jet structure. Since we are focused on exploring the impact of off-axis viewing angles in prompt emission properties rather than producing a realistic hardness-duration plane, we consider uniform distributions for all the parameters ($\gamma$, $\tau$, spectral parameters, $\rm \log_{10}\,{E_{iso}}, d_L^3,$ and $\cos \theta_v$). Our formalism does not consider the hardness-duration correlation, which requires imposing an empirical correlation between spectral and temporal parameters in the model. Since we do not include k-correction in our formalism, we have restricted $d_L$ to $500$~Mpc. For the same reason, we have taken a simplified approach to defining the duration as the extent of time the light curve is above a threshold value. A more accurate calculation will require including the detector response function and contributions of several noise components. In reality, factors such as extended emission can also affect the observed duration. Though we assume a single-pulse GRB, our conclusions also hold for bursts with complex light curves.

Given that the probability distribution of viewing angles $p(\theta_v) \propto \sin(\theta_v)$, off-axis jets are likely to be common. Previous studies have attempted to identify off-axis bursts, particularly following the GRB 170817A event, using its characteristics as distinguishing features \citep{grb150101b_2, grb150101b_1, gbmlike_170817_1, gbmlike_170817_2}. However, unambiguous identification can be challenging (also see \cite{xrf_030723, XRF020903_2, XRF020903, LLgrb_theory_1}). 

Extreme off-axis GRBs may remain undetectable, particularly if the jet profile exhibits a sharp cut-off, such as a top-hat jet. Our study demonstrates that these events can be detected if they are either nearby or luminous, and they are likely represented in the current population of detected bursts. Several low luminosity bursts have a soft spectrum \citep{low_lum_1, low_lum_2, low_lum5, low_lum_3}; they could, in fact, be off-axis events and may not be a part of a separate population. The rate of such events will also strongly depend on the GRB luminosity function, distribution of opening angles, and the jet structure. 

Soft-spectrum off-axis GRBs are more likely to be detected by the lower-energy instruments than by the higher-energy $\gamma$-ray monitors. It is possible that some of the X-ray transients and GRBs detected by the EP WXT (0.5 to 4 keV) and the SVOM/Eclairs (4 - 250 keV) are off-axis events. With these missions now operational, significant progress in GRB physics is anticipated as they provide a soft X-ray perspective on GRB prompt emission. The upcoming Indian mission Daksha \citep{daksha2} is also expected to make a landmark contribution in this field. Continued theoretical efforts to understand the off-axis prompt emission are crucial to complement these observations.

\begin{acknowledgments}
We thank the anonymous referee for their careful review and constructive feedback, particularly for pointing out the recent discoveries from EP and SVOM. LR acknowledges the grant MTR/2021/000830 from the Science and Engineering Research Board (SERB) of India. 
\end{acknowledgments}

\bibliography{sample631}{}
\bibliographystyle{aasjournal}
\end{document}